\documentstyle[prl,multicol,aps]{revtex}
\begin{document}

\draft
\title{Hyper-sparsity of the density matrix in a wavelet representation}
\author{S. Goedecker}
\address{ Max-Planck Institute for Solid State Research, Stuttgart, Germany}
\author{O. V. Ivanov}
\address{ P.N. Lebedev Physical Institute, Moscow, Russia }
\date{\today}
\maketitle

\begin{abstract}
O(N) methods are based on the decay properties of the density matrix 
in real space, an effect sometimes refered to as near-sightedness.
We show, that in addition to this near-sightedness in real space there is also 
a near-sightedness in Fourier space. Using a basis set with good localization 
properties in both real and Fourier space such as wavelets, one can exploit 
both localization properties to obtain a density 
matrix which exhibits additional sparseness properties compared to the 
scenario where one has a basis set with real space localization only.
We will call this additional sparsity hyper-sparsity. 
Taking advantage of this hyper-sparsity, 
it is possible to represent very large quantum mechanical 
systems in a highly compact way. This can be done both for insulating and 
metallic systems and for arbitrarily accurate basis sets. We expect that 
hyper-sparsity will pave the way for O(N) calculations of large systems requiring 
many basis functions per atom, such as Density Functional calculations.

\end{abstract}

\begin{multicols}{2}
\setcounter{collectmore}{5}
\raggedcolumns


Methods for the calculation of the electronic structure that exhibit 
linear scaling with respect to the size of the system, so-called 
O(N) methods~\cite{ON}, are an important topic in physics and chemistry. 
Only with this kind of algorithms 
it is possible to calculate the properties of very large systems 
that are relevant in many practical applications. O(N) methods 
have become practically a standard for tight binding calculations 
where the ratio of the number of basis functions to the number of 
electrons is of the order of 2. In highly accurate Density Functional type 
calculations, where the number of basis functions is very large 
compared to the number of electrons, these methods have on the other hand 
not been widely used up to now. 
This comes form the fact that O(N) methods are only efficient if the 
density matrix is very sparse. Large conventional basis sets lead however to 
density matrices that are not sufficiently sparse. The sparseness of the 
density matrix in connection with conventional localized basis sets is due 
to the real space decay properties of the density matrix, the so 
called real space near-sightedness~\cite{near} of the system. We will show 
how an additional near-sightedness in Fourier space can be used to 
make the density matrix significantly more sparse. 

For reasons of simplicity and in order to be able to do certain 
numerical calculations without any truncation we will concentrate in the following 
on the one dimensional case even though all the principles are applicable to the 
three dimensional case. For our investigations we use a 
test system that is characterized by a simple harmonic potential 
$-2 \sin( 2 \pi r)$ and we applied 
periodic boundary conditions. The valence band extends from roughly 0 to 4 atomic units 
and it is followed by a rather large gap of roughly 2 atomic units. By occupying 
each primitive cell that has a length of one atomic unit with one electron pair one thus 
obtains an insulator, while one obtains a metal by assigning for instance 2 electron 
pairs to each cell.

The density matrix $F$ in a real space representation can be modelled by 
\begin{equation}
F(r,r') = C \exp(\kappa r) \frac{\sin( \frac{\pi}{a} (r-r') )}{r-r'} 
\end{equation}
This is in principle only the asymptotic form, but 
is also a good approximation for small values of $r$ and $r'$. 
The decay constant $\kappa$ grows with increasing gap and is zero 
in a metallic system~\cite{sohrab}. The oscillation length $a$ is of the order of the 
average distance among the valence electrons 
in a metal and of the order of the interatomic spacing in an insulator.
$C$ is a normalization constant. Evidently $F(r,r')$ has both a typical length 
scale of $\frac{1}{\kappa}$ and a dominating Fourier component of $\frac{\pi}{a}$. 
The localization  in real space will be particularly good for strongly insulating systems 
whereas the localization around the typical Fourier component becomes better 
for decreasing gap sizes. It is well know that in most realistic materials 
$F(r,r')$ decays over several oscillation lengths indicating a combined localization 
in real and Fourier space. 

Let us now briefly review some essential facts from wavelet theory~\cite{daub}. 
For simplicity we will work with the standard Daubechies wavelet family~\cite{daub}.
There are two sets of fundamental functions, the scaling functions $\phi$ and the 
wavelets $\psi$. The scaling functions are essentially ordinary localized functions 
comparable for instance to Gaussians in many respects. Their advantage over 
Gaussians is however that they are orthogonal. To form a basis set one has to 
take all the integer translations $i$ of the scaling functions at a certain resolution 
level $k$
\begin{equation} \label{sf2ind}
\phi_{i}^{k}(x)  = \phi( 2^k x-i)
\end{equation}
Applying the unitary fast wavelet transform, one can transform the set of the 
scaling functions $\phi$ into the equivalent set of wavelets $\psi$.
The wavelet basis one 
obtains in this way consists now of translated wavelets at different resolution levels.
What is important in this context is that the scaling functions do not exhibit 
any localization in Fourier space whereas the wavelets do exhibit this feature.
The Fourier spectrum for the Daubechies wavelet of degree 12 (Figure~\ref{D12}) 
that was used in this work is shown in Figure~\ref{fourier}.
\begin{minipage}{3.375in}
   \begin{figure}[ht]
     \begin{center}
      \setlength{\unitlength}{1cm}
       \begin{picture}( 6.,5.0)           
        \put(-2.5,-.7){\includegraphics{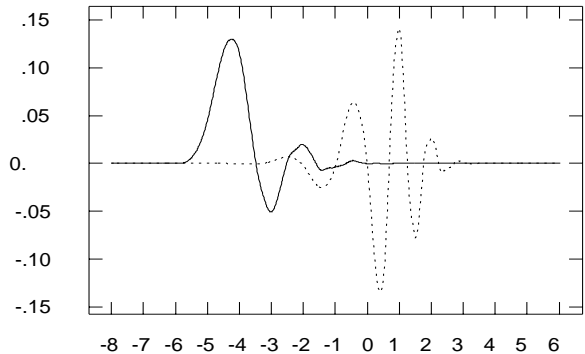}}   
       \end{picture}
       \caption{\label{D12} \it The Daubechies scaling function (solid line) 
                and wavelet (dashed line) used in this work.}
      \end{center}
     \end{figure}
\end{minipage}
\begin{minipage}{3.375in}
   \begin{figure}[ht]
     \begin{center}
      \setlength{\unitlength}{1cm}
       \begin{picture}( 6.,4.0)           
        \put(-2.0,-1.5){\includegraphics{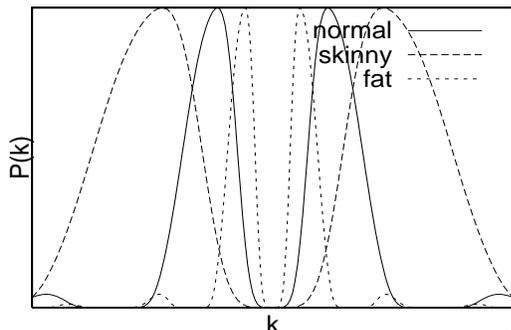}}   
       \end{picture}
       \caption{\label{fourier} \it The Fourier power spectrum $P(k)$ of 
                the Daubechies wavelet at three different resolution levels 
                denoted by skinny, normal and fat. 
                As one sees the peaks at the dominating Fourier components 
                are reasonably well separated from each other.}
      \end{center}
     \end{figure}
\end{minipage}

Let us now go over to O(N) context.
The most general O(N) methods are the ones that calculate the full density matrix. 
In these methods one calculates the matrix elements of $F$ 
with respect to a basis set $\chi_i$. 
\begin{equation} \label{fij}
F_{i,j} =  \int \int \chi_i(r) F(r,r') \chi_j(r') dr dr'
\end{equation}
Let us first discuss the case where the $\chi$'s are ordinary localized 
functions. In order to be able to do a fair evaluation of the advantages 
obtained by additional Fourier localization we will chose as our set of 
localized functions the set of scaling functions 
that is completely equivalent to the set of wavelets used in the comparison. 
It is obvious that the width of the scaling functions has to be less than the 
wavelength of the oscillatory part of $F$ in order to be an accurate basis.
This means that 
\begin{equation} \label{fscal}
F_{i,j} =  \int \int \phi^k_i(r) F(r,r') \phi^k_j(r') dr dr' \approx 
             F(R_i,R_j) 
\end{equation}
where $R_i$ is the position of the $i$-th scaling function. The function 
$F(0,r)$ for our test system is shown in Figure~\ref{frr}. 
\begin{minipage}{3.375in}
   \begin{figure}[ht]
     \begin{center}
      \setlength{\unitlength}{1cm}
       \begin{picture}( 6.,5.)           
        \put(-1.5,-.25){\includegraphics{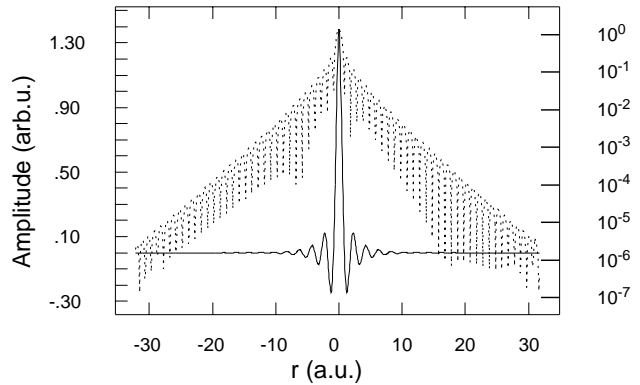}}   
       \end{picture}
       \caption{\label{frr} \it The function $F(0,r)$. Along the left 
                y-axis going with the solid line it is plotted on a normal scale and 
                along the right 
                y-axis going with the dashed line on a logarithmic scale.}
      \end{center}
     \end{figure}
\end{minipage}

Inspite of the exponential 
decay, one has to allow for a fairly large sub-volume of the whole 
64 atom system before one can truncate $F(0,r)$ without a large truncation error.
This means, that most of the elements of the density matrix have to be calculated 
and that therefore any traditional O(N) scheme will not be very efficient. 
The structure of the density matrix in this case is shown in Figure~\ref{dmscali}. 
Four scaling functions per atom were used, which gives a total energy that 
deviates by 4.e-3 a.u. per atom from the exact result in the limit of an 
infinite basis set.
\begin{minipage}{3.375in}
   \begin{figure}[ht]
     \begin{center}
      \setlength{\unitlength}{1cm}
       \begin{picture}( 6.,6.5)           
        \put(-2.5,-1.){\includegraphics{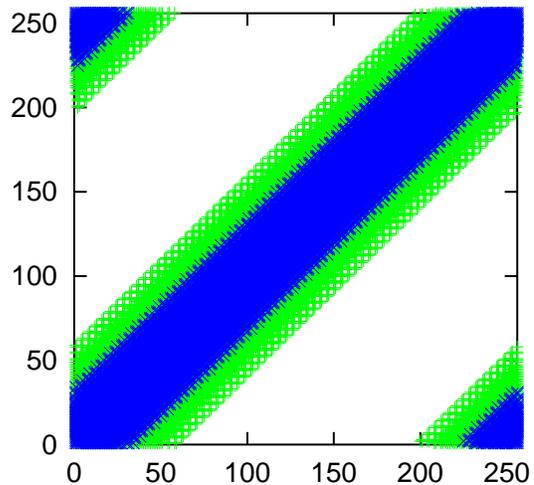}}   
       \end{picture}
       \caption{\label{dmscali} \it Structure of the density matrix in a scaling 
                function representation. Elements larger than 1.e-3 are denoted 
                by green areas elements larger than 1.e-2 by blue areas.}
      \end{center}
     \end{figure}
\end{minipage}

Let us now look at the same matrix in a wavelet basis, where we have to 
calculate matrix elements of the type 
\begin{equation} \label{fwvlt}
 \int \int \psi^k_i(r) F(r,r') \psi^{k'}_j(r') dr dr' 
\end{equation}
Note that in contrast to Equation~\ref{fscal} there are now matrix elements between 
wavelets at different resolution levels $k$ and $k'$, that have different dominating Fourier 
components. In accordance with the usual 
convention we order the wavelets at the different resolution levels 
such that the skinny wavelets follow the fat ones. So in our example, the indices 
129 to 256 refer to the skinniest wavelets, the indices 65 to 128 to the not quite 
so skinny wavelets etc.  Let us first look at the upper right 128 times 128 block 
in Figure~\ref{dmwvlti}. This block represents only matrix elements among the 
most skinny wavelets. From Figure~\ref{dmscali} we would expect 
to find elements larger than 1.e-2 and that the bandwidth 
is close to half of the size of the block. In reality all elements 
are smaller than 1.e-2, and the bandwidth is much less than the one in Figure~\ref{dmscali}.
This is the effect of the above postulated hyper-sparsity. If both basis functions 
in Equation~\ref{fwvlt} are skinny wavelets, the function $F(r,r')$ can not couple them 
since its dominating wave length is much larger than the dominating wave length 
of the two wavelets.  As a matter of fact any matrix element will be strongly 
damped unless all the three terms in Equation~\ref{fwvlt} have comparable wave lengths. 
Therefore also all the other fingers in Figure~\ref{dmscali} that represent 
coupling between wavelets of different stature (i.e. different $k$ and $k'$) 
have a small bandwidth. Because 
their spatial frequencies do not match, the decay for wavelets positioned a certain 
distance apart is much faster than one would expect from the decay behavior 
of the amplitude of $F(r,r')$ alone.
\begin{minipage}{3.375in}
   \begin{figure}[ht]
     \begin{center}
      \setlength{\unitlength}{1cm}
       \begin{picture}( 6.,6.5)           
        \put(-2.5,-1.){\includegraphics{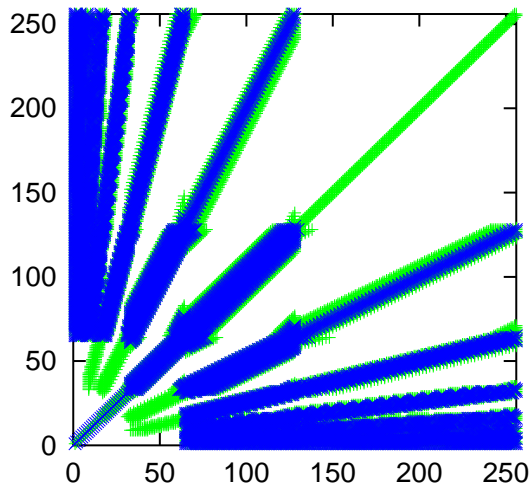}}   
       \end{picture}
       \caption{\label{dmwvlti} \it Structure of the density matrix in a wavelet 
                representation. The convention for the colors is the same as in 
                Figure~\ref{dmscali}. 
                The decay is so fast in this case that the green area around the blue area 
                is so thin to be hardly visible in some regions.}
      \end{center}
     \end{figure}
\end{minipage}

To quantitatively assess the advantages of the additional Fourier localization, 
compared with spatial localization only, let us plot the number of coefficients 
that are necessary to represent the density matrix with a certain error.
We define the error as the 2-norm of the difference between the exact 
and the truncated density matrix. As one can see from Figure~\ref{compress} 
the effect is dramatic. The accuracy obtained with the same number of 
coefficients with the wavelet representation is several orders of magnitude smaller 
than the one obtained with the scaling function representation. 
\begin{minipage}{3.375in}
   \begin{figure}[ht]
     \begin{center}
      \setlength{\unitlength}{1cm}
       \begin{picture}( 6.,5.0)           
        \put(-2.5,-1.){\includegraphics{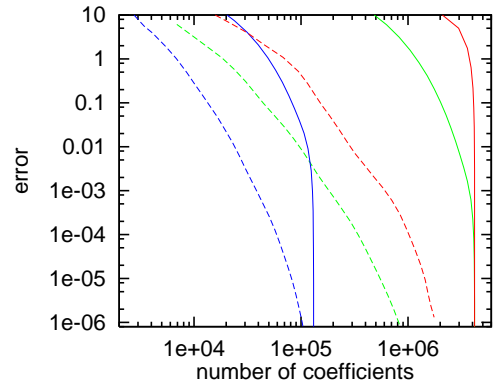}}   
       \end{picture}
       \caption{\label{compress} \it The error in the density matrix versus the 
                 size of the data set necessary for its representation. Solid lines correspond 
                 to a scaling function representation, dashed lines to a wavelet 
                 representation. The two red curves are for a metallic system, 
                 the two green curves for an insulator and the two blue curves
                 for an insulator where the density matrix was constructed indirectly 
                 via the Wannier functions.}
      \end{center}
     \end{figure}
\end{minipage}

In the case of an insulator at zero electronic temperature, the density matrix 
can compactly be represented in terms of Wannier functions. The Wannier 
functions are characterized by a spatial decay length and a dominant 
Fourier component as well. So the same principles apply and a wavelet representation 
is significantly more efficient than a scaling function representation 
(Figure~\ref{compress}). Since one needs only one Wannier function per electron 
pair instead of all the columns of the density matrix the prefactor in the Wannier 
representation is however smaller. 

Let us next consider the metallic systems. For the case of jellium, plane waves would 
obviously give the most compact representation, leading to a strictly diagonal 
density matrix. So the principle of spatial near-sightedness does not apply 
and the Fourier space localization of the basis functions 
will even be more important than the real space localization in the metallic case. 
As one sees from Figure~\ref{D12} 
our wavelets have a more pronounced spatial localization than Fourier localization. 
Other families of wavelet which have a more pronounced Fourier space localization 
might in this context be more efficient than the Daubechies wavelet used in these tests. 
The structure of the matrix is shown in Figure~\ref{dmwvltm} 
for the case of the wavelet representation of the test system containing 128 electron pairs 
and 512 basis functions. In the case of a scaling function representation 
all elements are larger than 1.e-2 and the corresponding matrix is therefore not shown. 
\begin{minipage}{3.375in}
   \begin{figure}[ht]
     \begin{center}
      \setlength{\unitlength}{1cm}
       \begin{picture}( 6.,6.5)           
        \put(-2.5,-1.){\includegraphics{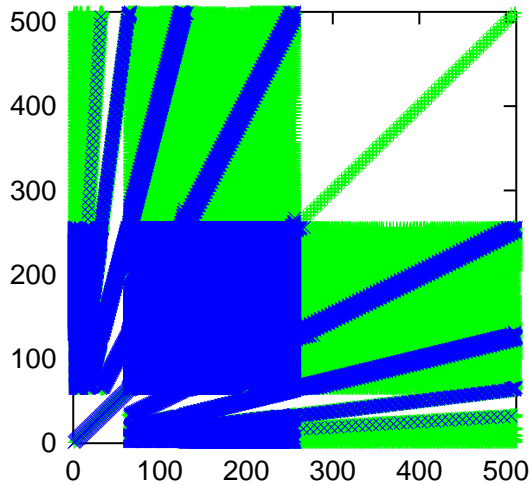}}   
       \end{picture}
       \caption{\label{dmwvltm} \it Same as Figure~\ref{dmwvlti} but for 
               a metallic system. }
      \end{center}
     \end{figure}
\end{minipage}

As expected, the quantitative evaluation in Figure~\ref{compress} shows that the savings 
of a wavelet representations compared to a scaling function representation 
are even more important in the metallic case than in the case of an insulator.  
In both representations more coefficients are however needed than in the insulating case. 
This comes from the fact that the decay behavior of matrix elements with moderate 
spatial frequency disparity is still determined by the slow spatial decay of $F(r,r')$. 
In addition to the finger structures there are therefore some full blocks 
in Figure~\ref{dmwvltm}. 

In the context of large systems, the scaling of the required number of 
basis functions with respect to system size is of course important. 
As is well known, even with basis sets which have real space localization only,  
this scaling is linear. Hence it is to be expected to be linear as well 
in the case of a combined spatial and Fourier localization. 
This assumption was clearly confirmed by our numerical experiments where we 
studied model systems containing between 16 and 128 "atoms". 
For metallic 
systems ordinary localized basis functions give a quadratic scaling unless one goes over to 
unrealistically large systems. 
Our numerical test for metallic systems containing between 16 and 128 "atoms" 
gave an exponent of 1.66. So the advantages of basis sets with frequency 
localization will grow with increasing system sizes.

Like in the majority of wavelet based calculations~\cite{arias,wei}, the 
basic computational kernels of the calculations presented here were implemented 
with respect to a scaling function 
basis. The final results were then transformed into a wavelet basis to examine the 
influence of truncation. In order to take full advantage of a truncated 
wavelet basis set, it would be 
necessary to do the whole calculation within the truncated wavelet basis. 
Matrix times vector multiplications are the basic step of several O(N) schemes and 
it has been demonstrated how they can be implemented with strictly linear 
scaling with respect to the number of non-zero coefficients~\cite{sgoi}. 
So the basic observation 
of this paper could directly be used to reduce the numerical effort in the same 
way as the number of necessary basis functions.

Based on the observation that the density matrix $F(r,r')$ is characterized by both 
a typical length scale and a dominating Fourier component, we have demonstrated that 
quantum mechanical systems exhibit not only a near-sightedness in 
real space but also a near-sightedness in Fourier space. A basis set 
that has combined localization in both real and Fourier space can therefore 
significantly reduce the number of data that are needed to describe it.
We expect that this phenomenon that we call hyper-sparsity will allow to 
overcome the present limitation of O(N) scheme to systems where the number of 
basis functions per atom is rather small. 
We thank J. Hutter, M. Parrinello, Anna Putrino and O. Jepsen for interesting discussions.

\end{multicols}

\begin{references}
\bibitem{ON}
 W. Yang, Phys. Rev. Lett. {\bf 66},1438  (1991) ;
 W. Yang and T-S. Lee, J. Chem. Phys. Rev. B {\bf 103},5674   (1995) ;
       S. Goedecker, M. Teter, Phys. Rev. B {\bf 51}, 9455 (1995) ;
 A. F. Voter, J. D. Kress and R. N. Silver, Phys. Rev. B {\bf 53}, 12733 (1996) ;
 S. Goedecker, J. of Comp. Phys. {\bf 118}, 261 (1995) ;
 X.-P. Li, W. Nunes and D. Vanderbilt, Phys. Rev. B{\bf 47},10891  (1993) ;
 M. S. Daw, Phys. Rev. B{\bf 47},10895  (1993) {\bf 6} 9153 (1994)  ;
 F. Mauri, G. Galli and R. Car, Phys. Rev. B {\bf 47},9973  (1993) ;
 P. Ordejon, D. Drabold, M. Grumbach and R. Martin, Phys. Rev. B {\bf 48}, 14646 (1993) ;
 J. Kim, F. Mauri and G. Galli, Phys. Rev. B {\bf 52},1640   (1995) ;
 A. Horsfield, A. Bratkovsky, D. Pettifor and M Aoki, Phys. Rev. B {\bf 53}, 1656 (1996)  ;
 E. Hernandez and M. Gillan, Phys. Rev. B {\bf 51}, 10157  (1995)
\bibitem{near} W. Kohn, Phys. Rev. Lett. {\bf 76},  (1996)
\bibitem{sohrab} S. Ismail-Beigi and T. Arias, submitted to Phys. Rev. Lett. 
\bibitem{daub} I. Daubechies, {\it ``Ten Lectures on Wavelets''}, SIAM, Philadelphia (1992)
\bibitem{arias}  K. Cho, T. Arias, J. Joannopoulos and P. Lam,
                Phys. Rev. Lett. {\bf 71}, 1808 (1993)
\bibitem{wei} S. Wei and M. Y. Chou, Phys. Rev. Lett. {\bf 76}, 2650 (1996)
\bibitem{sgoi}  S. Goedecker, O. Ivanov,  Sol. State Comm., {\bf 105}, 665 (1998) ; 
                  R. A. Lippert, T. Arias and A. Edelman, to appear in J. Comp. Physics ; 
                   T. Arias, to appear in Rev. of Mod. Phys. 


\end{references}
\end{document}